\documentclass[%
 reprint,
 amsmath,amssymb,
 aps,
]{revtex4-2}

\usepackage{graphicx}
\usepackage{dcolumn}
\usepackage{bm}

\usepackage[colorlinks=true, citecolor=blue, linkcolor=blue, urlcolor=blue]{hyperref}
\usepackage{amsmath}
\usepackage{subcaption}
\usepackage{caption}
\begin{document}

\preprint{APS/123-QED}

\title{Can spacetime torsion source an extremely red-tilted cosmological GW background?}

\author{Arko Bhaumik}
 \email{arkobhaumik12@gmail.com}
\author{Bhaswati Mandal}%
 \email{bhaswatimandaliitm92@gmail.com}
\affiliation{%
 Physics and Applied Mathematics Unit, Indian Statistical Institute, \\ 203, B.T. Road, Kolkata 700108, India
}%

\author{Soumitra SenGupta}
 \email{tpssg@iacs.res.in}
\affiliation{
 School of Physical Sciences, Indian Association for the Cultivation of Science \\
 2A \& 2B, Raja Subodh Chandra Mallick Road, Kolkata 700032, India
}%

\begin{abstract}
In the presence of spacetime torsion, any generic $f(R)$ model of gravity is conformally dual to a scalar-tensor theory augmented with a second rank antisymmetric massless degree of freedom. We investigate the stochastic gravitational wave background (SGWB) that may be sourced directly at the second order by such a torsional field, treated perturbatively during an epoch of canonical, single-field, slow-roll inflation. The resulting second-order induced SGWB, which dominates over the primary inflationary GW background at all scales, peaks only at ultra-low frequencies, and is found to be extremely red-tilted with an effective tensor spectral index $\alpha_{\rm T}\sim-6$ on matter-dominated scales. The signal is potentially within the reach of upcoming indirect GW probes on very large scales $k\lesssim10^{-2}\:\textrm{Mpc}^{-1}$, \emph{i.e.}, next-generation CMB experiments like the LiteBIRD. In the near future, observation of such a markedly red-tilted SGWB on CMB scales could hence provide a novel and unique clue in favour of torsional gravity during the inflationary era.
\end{abstract}

\maketitle


\underline{\bf Introduction:} A large-scale stochastic gravitational wave background (SGWB) remains one of the last major predictions of the inflationary paradigm that has remained elusive so far \cite{Guzzetti:2016mkm}. Latest observations of cosmic microwave background (CMB) anisotropies by Planck (Pl18) \cite{Planck:2018jri} and Bicep-Keck (BK15) \cite{BICEP2:2018kqh} have sharpened the upper bound on the inflationary tensor-to-scalar ratio to $r<0.036$ \cite{Campeti:2022vom} at the pivot scale $k_*=0.002$ Mpc\textsuperscript{-1}. Over the coming decades, next-generation gravitational wave (GW) detectors with superior sensitivity thresholds offer significantly better prospects for detecting a primordial SGWB, thus shedding light on both its size and spectral shape across a wide range of frequencies. 

The amplitude of an inflationary SGWB is directly tied to the inflationary energy scale, while its spectral tilt is a messenger of finer information, \emph{e.g.}, potential deviations from the simplest canonical, single field, slow-roll (CSFSR) scenario. The latter is well known for predicting a very weakly red-tilted spectrum of tensor perturbations \cite{Starobinsky:1979ty}. Thus, any detection of a very strongly red-tilted SGWB with a tensor spectral index $n_{\rm T}\lesssim-\mathcal{O}(1)$ on very large scales would be a game changer for ruling out large classes of inflationary models, and come as a breakthrough in terms of our understanding of the earliest moments of our Universe. At the same time, it would also strongly defy attempts at any non-inflationary causal explanation, thanks to the universal infrared (IR) scaling $n_{\rm T}\sim3$ for causally produced GW spectra \cite{Caprini:2009yp,Caprini:2009fx,Cai:2019cdl}.

As we show in this \emph{Letter}, an extremely red-tilted SGWB peaking on very large scales could interestingly serve as a smoking gun signature of torsion-augmented $f(R)$ gravity during inflation, while also allowing us to preserve the virtues of the CSFSR paradigm. Torsion arises naturally in the gravitational sector if the structure group acting on the tangent bundle of the spacetime manifold is identified with the Poincar\'{e} group, whose irreducible unitary representations are quantum fields labelled by mass and spin \cite{Hehl:1976kj,Grignani:1991nj,Arcos:2004tzt,Trautman:2006fp}. Within the framework of a generic $f(R)$ theory of gravity that includes torsion, transformation to the Einstein frame introduces two dynamical degrees of freedom $-$ a scalar field with a potential capable of driving CSFSR inflation, and a massless antisymmetric rank-2 tensor field that is minimally coupled at the leading order \cite{Kumar:2024bfc}. We demonstrate that the latter can source GWs at quadratic order over the course of inflation, resulting in an exceptionally red-tilted induced SGWB that dominates over the primary inflationary SGWB at all scales. The former could be detectable with upcoming space-based CMB experiments such as LiteBIRD \cite{LiteBIRD:2022cnt}, with a characteristic spectral index $\sim-6$ on CMB scales being a clear giveaway of its torsional origin. \\

\noindent
\underline{\bf Torsional field in Einstein frame:} The scalar-tensor representation of an $f(R)$ theory with torsion has been studied in \cite{Kumar:2024bfc}. The starting point is a diffeomorphism-invariant generic $f(R)$ action for gravity having the form
\begin{equation}
    S=\dfrac{1}{2\kappa^2}\int d^4x\sqrt{-g}\:f(R)\:,
\end{equation}
where $\kappa=M_{\rm Pl}^{-1}$ and all other symbols carry their usual meanings. This $f(R)$ action is known to be equivalent to
\begin{equation}
    S=\dfrac{1}{2\kappa^2}\int d^4x\sqrt{-g}\left[\Omega^2R-V\left(\Omega^2\right)\right]\:,
\end{equation}
where $f'(A)=\Omega^2$ and $V(\Omega^2)=A(\Omega^2)\Omega^2-f(A(\Omega^2))$.
In the standard scenario, one can proceed to the Einstein frame by conformally transforming $g_{\mu\nu}\to\widetilde{g}_{\mu\nu}=\Omega(x)^2g_{\mu\nu}$, which ``frees'' the Ricci scalar and causes a scalar degree of freedom to emerge \cite{Sotiriou:2008rp}.

However, since the Christoffel connection ($\Gamma$) is no longer symmetric in twisted spacetime, it decomposes as $\Gamma^{\lambda}_{\mu\nu}=\bar{\Gamma}^{\lambda}_{\mu\nu}-K^\lambda_{\:\mu\nu}$, where the first term is its symmetric component and the second term is the antisymmetric contorsion tensor. This entails non-trivial additive contributions from $K^\lambda_{\mu\nu}$ in the Ricci tensor. Consequently, one finds that the Einstein frame action ($S_E$) involves a scalar degree of freedom ($\phi$) as well as a rank-2 antisymmetric tensor field ($Z_{\mu\nu}$). The final canonical form of $S_E$ may be expressed as follows \cite{Kumar:2024bfc}:
\begin{align} \label{eq:SEinst}
    &S_E=\int d^4x\sqrt{-g}\bigg[\dfrac{R}{2\kappa^2}-\dfrac{1}{2}g^{\alpha\beta}\partial_\alpha\phi\partial_\beta\phi-V(\phi) \nonumber \\
    &-\dfrac{1}{2}\nabla_{[\alpha}Z_{\mu\nu]}\nabla^{[\alpha}Z^{\mu\nu]} +{\kappa}\sqrt{\dfrac{6}{7}}g^{\alpha\gamma}g^{\beta\epsilon}g^{\lambda\delta}\partial_{[\alpha}\phi Z_{\epsilon\delta]}\partial_{[\beta}Z_{\lambda\gamma]} \nonumber \\
    &-\dfrac{3\kappa^2}{7}g^{\alpha\gamma}g^{\beta\epsilon}g^{\lambda\delta}\partial_{[\alpha}\phi Z_{\epsilon\delta]}\partial_{[\beta}\phi Z_{\lambda\gamma]}\bigg]\:,
\end{align}
where all quantities are defined with reference to the Einstein frame. While the first line contains the terms typical to $f(R)$ theories without torsion, the rest emerge uniquely from the torsional sector. In particular, besides the $\mathcal{O}(Z^2)$ kinetic term of the massless torsional field, there are $\mathcal{O}(\phi Z^2)$ and $\mathcal{O}(\phi^2Z^2)$ interaction terms which are Planck-suppressed. In the present work, we neglect these higher-order terms. \\

\noindent
\underline{\bf First order dynamics of $Z_{\mu\nu}$:} We now specialize to the case of a homogeneous and isotropic, flat Friedmann-Lema\^{i}tre-Robertson-Walker (FLRW) metric $ds^2=-dt^2+a(t)^2\boldsymbol{dx}_3^2$ as the cosmological background. We assume the torsional field to be energetically subdominant compared to the scalar $\phi$, which is advantageous to identify with the inflaton. Note that all information of the functional form of $f(R)$ is distilled into the scalar potential $V(\phi)$, while the torsional sector remains independent of it. Hence, one may assume CSFSR inflation of the background driven by the scalar rolling down a suitably constructed $V(\phi)$, with $Z_{\mu\nu}$ invoked perturbatively on top. The background equations thus remain
\begin{equation}
    H^2=\dfrac{\kappa^2}{3}\left[\dfrac{1}{2}\dot\phi^2+V(\phi)\right]\:\:,\:\:\:\ddot\phi+3H\dot\phi+\dfrac{\partial V}{\partial\phi}=0\:,
\end{equation}
where the overdots denote derivatives with respect to $t$, and $H(t)=\dot{a}(t)/a(t)$. As in the standard scenario, this is sufficient to ensure a quasi-de Sitter (dS) expansion of the background, governed by the specific choice of $f(R)$ that may lead to a suitably flat $V(\phi)$. Invoking $Z_{\mu\nu}(\vec{x},t)$ perturbatively on top of this background and neglecting the higher-order couplings of \eqref{eq:SEinst}, its EoM may be written component-wise as
\begin{gather}
    \nabla^2Z_{i0}+\partial_\ell\left(\dot{Z}_{\ell i}+\partial_iZ_{0\ell}\right)=0\:, \label{eq:Zi0eqn} \\
    \left(\dfrac{d}{dt}+3H\right)\left(\dot{Z}_{ij}+\partial_i Z_{j0}-\partial_j Z_{i0}\right) \nonumber \\
    -\dfrac{1}{a^2}\left[\nabla^2Z_{ij}+\partial_\ell\left(\partial_iZ_{j\ell}-\partial_jZ_{i\ell}\right)\right]=0\:. \label{eq:Zijeqn}
\end{gather}
To simplify this system, we impose the following pair of constraints on $Z_{\mu\nu}$: (i) the time-space components of $Z_{\mu\nu}$ are identically zero, \emph{i.e.}, $Z_{i0}=0\:\forall\:i\in\{1,2,3\}$, and (ii) the space-space components of $Z_{\mu\nu}$ are transverse, \emph{i.e.}, $\partial_iZ_{ij}=0\:\forall\:j\in\{1,2,3\}$. Such conditions may be physically motivated based on gauge symmetries, if, for example, the torsional field has a string-theoretic interpretation \cite{KRorig,Berche:2022sel,Capanelli:2023uwv}. The constraints above leave $6-3-2=1$ propagating transverse degree of freedom in $Z_{\mu\nu}$, which is hence dual to a massless scalar. With \eqref{eq:Zi0eqn} rendered superfluous, \eqref{eq:Zijeqn} is given in $k$-space in terms of $\mathcal{Z}_{ij}=aZ_{ij}$ and the conformal time $d\eta=dt/a(t)$ as
\begin{equation} \label{eq:Zeqnfin}
    \mathcal{Z}''_{ij}(\boldsymbol{k},\eta)+\left(\dfrac{k^2}{2}-\dfrac{2}{\eta^2}\right)\mathcal{Z}_{ij}(\boldsymbol{k},\eta)=0\:,
\end{equation}
where $a(\eta)=-(\eta H)^{-1}$ has been used, as consistent with an inflating dS background. The steps to quantize $\mathcal{Z}_{ij}$ are outlined in Appendix \ref{sec:appA}, following which the Bunch-Davies (BD) normalized solution for the mode function of \eqref{eq:Zeqnfin} is given by
\begin{equation} \label{eq:zmodesol}
    \mathcal{Z}_k(\eta)=\dfrac{2^{\frac{1}{4}}}{\sqrt{k}}\dfrac{e^{-\frac{ik\eta}{\sqrt{2}}}}{k\eta}\left(1+\dfrac{ik\eta}{\sqrt{2}}\right)\:.
\end{equation}
In subsequent calculations, we have switched back to $Z_k(\eta)=a^{-1}\mathcal{Z}_k(\eta)$. Based on \eqref{eq:zmodesol}, the superhorizon behavior of $Z_k(\eta)$ can be approximated as $Z_k^{(0)}=Z_k(\eta)\lvert_{|k\eta|\ll1}\:\approx -2^{\frac{1}{4}}H/k^{\frac{3}{2}}$, which is constant. Thus, one may introduce an inflationary transfer function $T_k(\eta)$ such that $Z_k(\eta)=Z_k^{(0)}T_k(\eta)$, yielding 
\begin{equation} \label{eq:inftransf}
    T_k(\eta)=e^{-\frac{ik\eta}{\sqrt{2}}}\left(1+\dfrac{ik\eta}{\sqrt{2}}\right)\:,
\end{equation}
which is used in our subsequent calculations. \\

\noindent
\underline{\bf Torsion-sourced SGWB:} We now consider a spatially perturbed flat FLRW metric with $g_{ij}=a^2(\delta_{ij}+h_{ij})$, where the tensor perturbations $h_{ij}$ are identified with GWs propagating on the expanding background. In presence of an anisotropic stress source, they follow the EoM
\begin{equation} \label{eq:heqn}
    h_{ij}''+2\mathcal{H}h_{ij}'+k^2 h_{ij}=\kappa^2a^2T_{ij}^{(T)}\:,
\end{equation}
with $T_{ij}^{(T)}\equiv\hat{\mathcal{T}}^{\ell m}_{ij}T_{\ell m}$ being the spatial transverse-traceless (TT) projection of the source energy-momentum tensor. In our case, the latter is furnished by the $\mathcal{O}(Z^2)$ kinetic term of \eqref{eq:SEinst}, which results in
\begin{equation} \label{eq:ztij}
    T_{ij}^{(T)}[Z]=\dfrac{2}{3a(\eta)^4}\left[Z_{\ell i}'Z_{\ell j}'+\partial_\ell Z_{mi}\left(\partial_\ell Z_{mj}+\partial_m Z_{j\ell}\right)\right]\:.
\end{equation}
Using the helicity-basis decomposition $h_{ij}(\boldsymbol{k},\eta)=\sum\limits_{\lambda=\pm}h_\lambda(\boldsymbol{k},\eta)\Pi^{ij}_\lambda(\boldsymbol{k})$, where $\Pi^{ij}_{\pm}(\boldsymbol{k})=e_{\mp}^i(\boldsymbol{k})e_{\mp}^j(\boldsymbol{k})/\sqrt{2}$ is written in terms of orthonormal polarization vectors, \eqref{eq:heqn} leads to the mode equation
\begin{equation} \label{eq:hmodeeqn}
    h_{\lambda}''+2\mathcal{H}h_{\lambda}'+k^2h_{\lambda}=\kappa^2a^2\Pi^{ij}_{\lambda}(\boldsymbol{k})\widetilde{T}_{ij}^{(T)}[Z]=S_\lambda(\boldsymbol{k},\eta)\:,
\end{equation}
where $\widetilde{T}$ denotes the Fourier transform of $T$ from \eqref{eq:ztij}. Using convolutions for the quadratic terms, the source function can be expressed in $k$-space as
\begin{align} \label{eq:Sfunc}
    S_\lambda(\boldsymbol{k},\eta)=&\dfrac{2\kappa^2}{3a(\eta)^2}\Pi^{ij}_\lambda(\boldsymbol{k})\int\dfrac{d^3k'}{(2\pi)^3}\bigg[Z_{\ell i}(\boldsymbol{k}',\eta)Z_{\ell j}(\boldsymbol{k}-\boldsymbol{k}',\eta) \nonumber \\
    &-k'_\ell(k_\ell-k'_\ell) Z_{mi}(\boldsymbol{k}',\eta)Z_{mj}(\boldsymbol{k}-\boldsymbol{k}',\eta) \nonumber \\
    &-k'_\ell(k_m-k'_m)Z_{mi}(\boldsymbol{k}',\eta)Z_{j\ell}(\boldsymbol{k}-\boldsymbol{k}',\eta)\bigg]\:.
\end{align}
The tensor mode function may then be written as
\begin{equation}
    h_\lambda(\boldsymbol{k},\eta)=\int\limits_{\eta_{\rm in}}^\eta d\tilde{\eta}\:G_k(\eta,\tilde{\eta})S_\lambda(\boldsymbol{k},\tilde{\eta})\:,
\end{equation}
where the Green's function $G_{k}(\eta,\tilde{\eta})$ for \eqref{eq:hmodeeqn} in dS background reads
\begin{eqnarray}
    G_k(\eta,\tilde{\eta})&&=\dfrac{1}{k^3\tilde{\eta}^2}\left[\left(k\tilde{\eta}-k\eta\right)\cos(k\eta-k\tilde{\eta}) \right. \nonumber \\
    &&\left. +\left(k^2\eta\tilde{\eta}+1\right)\sin(k\eta-k\tilde{\eta})\right]\Theta(\eta-\tilde{\eta})\:,
\end{eqnarray}
with $\Theta(x)$ being the Heaviside step function \cite{green_eqn}. This subsequently enables one to construct the equal-time two-point tensor correlator, whose full analytic expression is provided in Appendix \ref{sec:appB}. Explicit computation reveals $\langle h_+h_+\rangle=\langle h_-h_-\rangle$ and $\langle h_+h_-\rangle =\langle h_-h_+\rangle=0$. Hence, the corresponding $Z$-induced tensor power spectrum can be defined as
\begin{eqnarray} \label{eq:tenspowspec}
    \langle h_+(\boldsymbol{k},\eta)h_+(\boldsymbol{\tilde{k}},\eta)\rangle=\dfrac{\pi^2}{k^3}\delta^{(3)}(\boldsymbol{k}+\boldsymbol{\tilde{k}})\:\mathcal{P}_h^{(Z)}(k,\eta)\:.
\end{eqnarray}

With \eqref{eq:tenspowspec} at hand, we are ready to compute the tensor power spectrum numerically. First, we transform to the dimensionless variables $v=k'/k$ and $u=|\boldsymbol{k}-\boldsymbol{k}'|/k$. While integrating over $k'$, we assume an ultraviolet (UV) cutoff $k_{\rm max}=k_i=a_iH$, with $a_i$ being the scale factor at the end of inflation, approximated as \cite{ai_eqn}
\begin{equation}
    a_i=\left(0.9\times10^{29}\right)^{-1}\left(\dfrac{H}{10^{-5}M_{\rm Pl}}\right)^{-1/2}\:
\end{equation}
for instantaneous reheating. We also impose an IR cutoff corresponding to the largest observable scale corresponding to the present horizon size, \emph{i.e.}, $k_{\rm min}=H_0\sim2\times10^{-4}\:\rm Mpc^{-1}$. The $u$-integral then runs between $\max\left[k_{\rm min}/k_1,|1-v|\right]$ and $\min\left[k_{\rm max}/k_1,1+v\right]$, and the $v$-integral between $0$ and $k_{\rm max}/k_1$. For time integration, we accordingly choose $\eta_{\rm in}=-\eta_0$ (with $\eta_0$ being the conformal time at present) and the final time $\eta_{\rm end}=-k_i^{-1}$, \emph{i.e.}, when the longest and the shortest observable modes exited the horizon during inflation. This spans the entire duration of physical inflation, and encapsulates the contribution from the full range of scales therein.

Having thus computed the GW power spectrum at the end of inflation ($\eta_{\rm end}$), the GW spectral abundance at present ($\eta_0$) is finally calculated as $\Omega_{\rm GW}(k,\eta_0)=T_h\left(k;\eta_0,\eta_{\rm end}\right)^2\mathcal{P}_h^{(Z)}(k,\eta_{\rm end})$\:, where $T_h$ denotes the sub-horizon transfer function describing the evolution of tensor modes in the post-inflationary Universe (see Appendix \ref{sec:appB}). \\

\begin{figure*}[t!]
    \centering
    \begin{subfigure}[b]{0.5\textwidth}
        \centering
        \includegraphics[width=\textwidth]{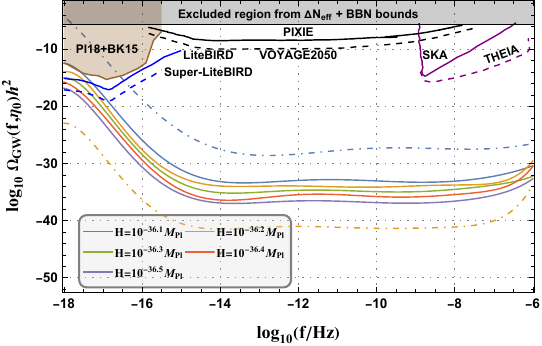}
    \end{subfigure}%
    ~ 
    \begin{subfigure}[b]{0.5\textwidth}
        \centering
        \includegraphics[width=\textwidth]{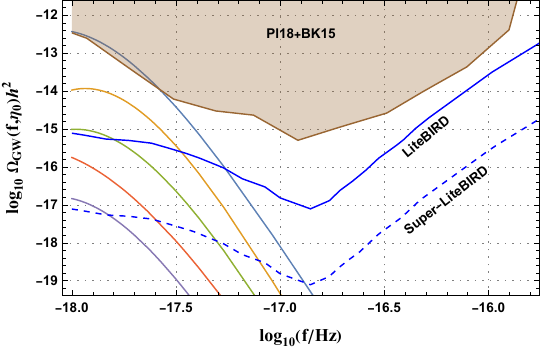}
    \end{subfigure}
    \caption{Present day spectral abundance of the torsion-sourced SGWB, consistent with currently available upper bounds from BBN+$\Delta N_{\rm eff}$ constraints \cite{Yeh:2022heq} and CMB measurements of Pl18+BK15 \cite{Clarke:2020bil}. We focus on a few viable values of the inflationary Hubble parameter ($H$) for which the prominently red-tilted SGWB should be detectable by upcoming CMB probes, \emph{e.g.}, the LiteBIRD and its hypothetical successor(s). Two more curves for $H=10^{-35.5}M_{\rm Pl}$ (blue dot-dashed) and $H=10^{-37}M_{\rm Pl}$ (orange dot-dashed) have been included for the purpose of demonstration, with the former overshooting the admissible upper bound and the latter falling below the detectable threshold. The projected sensitivity curves of a few other proposed future instruments have also been shown for reference \cite{2011JCAP...07..025K,Chluba:2019nxa,SKA:2018ckk,2018FrASS...5...11V}. The right panel displays a zoomed-in region from the left panel highlighting detectability at CMB scales.}
    \label{fig:cmbplot}
\end{figure*}

\begin{figure}[t!]
    \centering
    \includegraphics[width=0.5\textwidth]{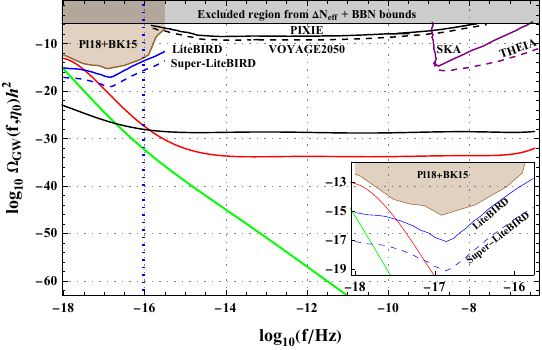}
    \caption{Comparison of the spectral shape of the torsion-sourced SGWB (red), with those of the primary inflationary SGWB for $n_{\rm T}=-6$ (green) and $n_{\rm T}=0$ (black), for $H=10^{-36.2}M_{\rm Pl}$. The amplitudes of the primary SGWB are scaled up overall by $10^{50}$. Across MD scales towards the left of the matter-radiation equality scale (vertical blue-dotted line), the torsion-sourced spectrum visibly has an effective tensor spectral index $\alpha_{\rm T}\sim-6$.}
    \label{fig:primcomp}
\end{figure}

\noindent
\underline{\bf Results and Discussion:} In Fig. \ref{fig:cmbplot}, we show the spectral profile envelopes of the torsion-sourced SGWB for a few representative values of the inflationary Hubble parameter ($H$), which is the only governing parameter in our scenario. We particularly focus on the parameter region which produces an SGWB consistent with presently available datasets and their resultant upper bounds on the SGWB strength, while being of interest to future-generation CMB-scale detectors where the signal strength peaks. Remarkably, the avoidance of GW overproduction by the torsional field requires $H\lesssim10^{-36}M_{\rm Pl}$, which is significantly smaller than the currently available data-based upper bound $H\lesssim10^{-5}M_{\rm Pl}$. However, this is not very problematic, as the theoretical lower bound is much smaller and lies around $H\gtrsim10^{-44}M_{\rm Pl}$ based on the Big Bang Nucleosynthesis (BBN) energy scale $\sim1$ MeV \cite{Fields:2019pfx}. As a direct result of such lowered energy scale of inflation, the primary inflationary GW background remains negligibly small compared to the torsion-sourced SGWB, with a maximum amplitude of $\sim10^{-6}(k_{\rm eq}/k_{\rm min})^2(H/M_{\rm Pl})^2\lesssim10^{-74}$ during the matter-dominated (MD) era, where $k_{\rm eq}\sim10^{-2}\:\textrm{Mpc}^{-1}$ denotes the scale of matter-radiation equality.

The SGWB amplitude peaks close to $f\sim10^{-18}$ Hz and quickly decays at higher frequencies, hence falling beyond the scope of detectability at all but the largest scales. In analogy with the power-law primary tensor power spectrum $\mathcal{P}_{h}^{\rm (pr)}(k)=A_{\rm T}(k/k_*)^{n_{\rm T}}$, one may define an effective tensor spectral index $\alpha_{\rm T}$ such that $\mathcal{P}_{\rm T}^{(Z)}(k)\propto k^{\alpha_{\rm T}}$ for scales re-entering in the MD era. The simplest way to estimate $\alpha_{\rm T}$ is to compare with the tilt of the spectral abundance of primary inflationary GWs during the MD era, whose spectral shape for $n_{\rm T}=-6$ is shown in Fig. \ref{fig:primcomp}. A direct comparison between the spectral shapes of the primary spectrum and the torsion-sourced spectrum therefore yields an approximate value of $\alpha_{\rm T}\sim-6$, confirming that the torsion-sourced SGWB is strikingly red-tilted. In hindsight, this is intuitive as the torsional source is a second-rank antisymmetric tensor field, which implies an overall $a^{-4}$ dilution of its energy density with cosmic expansion as seen in \eqref{eq:ztij}. Consequently, GW production is expected to have happened predominantly during the earliest period of inflation, when the longest observable comoving modes were leaving the horizon. \\

\noindent
\underline{\bf Conclusion:} While torsion is an intrinsic property of spacetime geometry, its role in present-day gravitational interactions is observed to be negligibly small compared to that of curvature. However, this need not always have been the case in the past. In this \emph{Letter}, we have, for the first time, considered a torsional field to be the direct physical source of a cosmological GW background generated during inflation. The scenario follows simply from the inclusion of spacetime torsion in any generic model of $f(R)$ gravity, which, in the Einstein frame, quite naturally furnishes both a scalar degree of freedom that may drive CSFSR inflation for some suitable choice of $f(R)$, and an antisymmetric rank-2 tensor field which is capable of sourcing GWs perturbatively at the second order. Importantly, the viability of the whole setup apparently requires an exceptionally lowered energy scale of inflation ($H\lesssim10^{-36}M_{\rm Pl}$), which is the only parameter in the model under consideration, to ensure that GWs are not overproduced by the torsional field. The same condition also guarantees that the secondary SGWB dominates over the primary inflationary GW background at all scales. Most significantly, we find that the scenario leads to an extremely red-tilted induced SGWB peaking within the sensitivity thresholds of upcoming CMB missions, \emph{e.g.} the LiteBIRD satellite. Based on our results, detecting such a GW background with a characteristic spectral tilt $\alpha_{\rm T}\sim-6$ on CMB scales in the near future may provide us with a telltale signature of spacetime torsion from the earliest moments of our Universe. \\

\noindent
\underline{\bf Acknowledgments:} Authors thank Rahul Shah and Supratik Pal for fruitful discussions. AB thanks CSIR for financial support through Senior Research Fellowship (File no. 09/0093(13641)/2022-EMR-I).

\appendix

\section{Quantization of $\mathcal{Z}_{ij}$} \label{sec:appA}

The canonically normalized Minkowskian action for \eqref{eq:Zeqnfin} is given by
\begin{equation} \label{eq:Sact}
    S_{\mathcal{Z}\mathcal{Z}}=\dfrac{1}{2}\int d^3k\:d\eta\left(\mathcal{Z}_{ij}'^2-\dfrac{k^2}{2}\mathcal{Z}_{ij}^2-\dfrac{a''}{a}\mathcal{Z}_{ij}^2\right)\:.
\end{equation}
Equipped with the conjugate momentum $\Pi^{(\mathcal{Z})}_{ij}(\boldsymbol{k},\eta)= \dfrac{\delta S_{\mathcal{Z}\mathcal{Z}}}{\delta \mathcal{Z}_{ij}'(\boldsymbol{k},\eta)}=\mathcal{Z}_{ij}'(\boldsymbol{k},\eta)$, we introduce the mode expansion 
\begin{equation}
    \hat{\mathcal{Z}}_{ij}(\boldsymbol{k},\eta)=\sum\limits_{\lambda=\pm}\varepsilon_{ij}^{(\lambda)}(\boldsymbol{k})\left[\mathcal{Z}_k(\eta)\hat{c}_{\boldsymbol{k},\lambda}+\mathcal{Z}^*_k(\eta)\hat{c}^\dagger_{-\boldsymbol{k},\lambda}\right]\:,
\end{equation}
where the creation and annihilation operators satisfy $\left[\hat{c}_{\boldsymbol{k},\lambda},\hat{c}^\dagger_{\boldsymbol{k}',\lambda'}\right]=(2\pi)^3\delta_{\lambda\lambda'}\delta^{(3)}(\boldsymbol{k}-\boldsymbol{k}')$, and the transverse basis tensors are constructed based on the antisymmetrization requirement as $\varepsilon_{ij}^\lambda(\boldsymbol{k})=\sqrt{2}e_{-\lambda}^{[i}(\boldsymbol{k})e_{-\lambda}^{j]}(\boldsymbol{k})$. We then use the canonical field commutation relation
\begin{equation}
    \left[\mathcal{Z}_{ij}(\boldsymbol{k},\eta),\Pi^{(\mathcal{Z})}_{pq}(\boldsymbol{k}',\eta)\right]=\dfrac{i}{2}\left(\delta_{ip}\delta_{jq}-\delta_{iq}\delta_{jp}\right)\delta^{(3)}(\boldsymbol{k}-\boldsymbol{k}')
\end{equation}
together with the $\left[c,c^\dagger\right]$ commutator to arrive at the Wronskian condition for the mode functions given by
\begin{equation} \label{eq:wronskian}
    \mathcal{Z}_k(\eta)\mathcal{Z}_k^{*'}(\eta)-\mathcal{Z}_k^*(\eta)\mathcal{Z}_k^{'}(\eta)=i\:,
\end{equation}
which makes sense as \eqref{eq:Sact} is canonically normalized. This Wronskian serves as the requisite normalization criterion for the mode functions $\mathcal{Z}_k(\eta)$, which are subsequently obtained by solving \eqref{eq:Zeqnfin}. 

\section{Analytic expressions for the GW two-point function} \label{sec:appB}

As the source function given in \eqref{eq:Sfunc} contains three distinct terms, the overall $\langle S_\lambda(\boldsymbol{k},\eta_1)S_{\lambda'}(\boldsymbol{\tilde{k}},\eta_2)\rangle$ correlator contains a total of nine terms. Thereafter, integrating over $\eta_1$ and $\eta_2$ across identical finite intervals $\left[\eta_{\rm in},\eta_{\rm end}\right]$ leads to the following terms in the induced GW two-point correlation function evaluated at the end of inflation:

\begin{widetext}
\allowdisplaybreaks
\begin{align}
    &\langle h_\lambda(\boldsymbol{k},\eta_{\rm end})h_{\lambda'}(\boldsymbol{\tilde{k}},\eta_{\rm end})\rangle_{(1)}=\dfrac{16\kappa^4H^4}{9}\delta^{(3)}(\boldsymbol{k}+\boldsymbol{\tilde{k}})\int\dfrac{d^3k'}{k'^3|\boldsymbol{k}-\boldsymbol{k'}|^3}\Pi_\lambda^{ij}(\boldsymbol{k})\Pi_{\lambda'}^{pq}(\boldsymbol{k})^* \nonumber \\
    &\times\sum\limits_{\lambda_1,\lambda_2=\pm}\varepsilon_{\ell i}^{(\lambda_1)}(\boldsymbol{k'})\varepsilon_{mp}^{(\lambda_1)}(\boldsymbol{k'})^*\varepsilon_{\ell j}^{(\lambda_2)}(\boldsymbol{k}-\boldsymbol{k'})\varepsilon_{mq}^{(\lambda_2)}(\boldsymbol{k}-\boldsymbol{k'})^*\Bigg\lvert\int\limits_{\eta_{\rm in}}^{\eta_{\rm end}}\dfrac{d\eta_1}{a(\eta_1)^2}G_k(\eta_{\rm end},\eta_1)f_Z(k',|\boldsymbol{k}-\boldsymbol{k'}|,\eta_1)\Bigg\rvert^2\:,
\end{align}
\begin{align}
    &\langle h_\lambda(\boldsymbol{k},\eta_{\rm end})h_{\lambda'}(\boldsymbol{\tilde{k}},\eta_{\rm end})\rangle_{(2)+(3)}=-\dfrac{32\kappa^4H^4}{9}\delta^{(3)}(\boldsymbol{k}+\boldsymbol{\tilde{k}})\int\dfrac{d^3k'}{k'^3|\boldsymbol{k}-\boldsymbol{k'}|^3}\boldsymbol{k'}.(\boldsymbol{k}-\boldsymbol{k'})\nonumber \\
    &\times\Re\Bigg[\Pi_\lambda^{ij}(\boldsymbol{k})\Pi_{\lambda'}^{pq}(\boldsymbol{k})^*\times\sum\limits_{\lambda_1,\lambda_2=\pm}\varepsilon_{m i}^{(\lambda_1)}(\boldsymbol{k'})\varepsilon_{\ell p}^{(\lambda_1)}(\boldsymbol{k'})^*\varepsilon_{m j}^{(\lambda_2)}(\boldsymbol{k}-\boldsymbol{k'})\varepsilon_{\ell q}^{(\lambda_2)}(\boldsymbol{k}-\boldsymbol{k'})^* \nonumber \\
    &\times\int\limits_{\eta_{\rm in}}^{\eta_{\rm end}}\dfrac{d\eta_1}{a(\eta_1)^2}G_k(\eta_{\rm end},\eta_1)f_Z(k',|\boldsymbol{k}-\boldsymbol{k'}|,\eta_1)\int\limits_{\eta_{\rm in}}^{\eta_{\rm end}}\dfrac{d\eta_2}{a(\eta_2)^2}G_k(\eta_{\rm end},\eta_2)\xi_Z(k',|\boldsymbol{k}-\boldsymbol{k'}|,\eta_2)^*\Bigg]\:,
\end{align}
\begin{align}
    &\langle h_\lambda(\boldsymbol{k},\eta_{\rm end})h_{\lambda'}(\boldsymbol{\tilde{k}},\eta_{\rm end})\rangle_{(4)+(5)}=-\dfrac{32\kappa^4H^4}{9}\delta^{(3)}(\boldsymbol{k}+\boldsymbol{\tilde{k}})\int\dfrac{d^3k'}{k'^3|\boldsymbol{k}-\boldsymbol{k'}|^3}k'_\ell.(k_m-k_m')\nonumber \\
    &\times\Re\Bigg[\Pi_\lambda^{ij}(\boldsymbol{k})\Pi_{\lambda'}^{pq}(\boldsymbol{k})^*\times\sum\limits_{\lambda_1,\lambda_2=\pm}\varepsilon_{m i}^{(\lambda_1)}(\boldsymbol{k'})\varepsilon_{\alpha p}^{(\lambda_1)}(\boldsymbol{k'})^*\varepsilon_{j\ell}^{(\lambda_2)}(\boldsymbol{k}-\boldsymbol{k'})\varepsilon_{\alpha q}^{(\lambda_2)}(\boldsymbol{k}-\boldsymbol{k'})^* \nonumber \\
    &\times\int\limits_{\eta_{\rm in}}^{\eta_{\rm end}}\dfrac{d\eta_1}{a(\eta_1)^2}G_k(\eta_{\rm end},\eta_1)\xi_Z(k',|\boldsymbol{k}-\boldsymbol{k'}|,\eta_1)\int\limits_{\eta_{\rm in}}^{\eta_{\rm end}}\dfrac{d\eta_2}{a(\eta_2)^2}G_k(\eta_{\rm end},\eta_2)f_Z(k',|\boldsymbol{k}-\boldsymbol{k'}|,\eta_2)^*\Bigg]\:,
\end{align}
\begin{align}
    &\langle h_\lambda(\boldsymbol{k},\eta_{\rm end})h_{\lambda'}(\boldsymbol{\tilde{k}},\eta_{\rm end})\rangle_{(6)+(7)+(8)+(9)}=\dfrac{16\kappa^4H^4}{9}\delta^{(3)}(\boldsymbol{k}+\boldsymbol{\tilde{k}})\int\dfrac{d^3k'}{k'^3|\boldsymbol{k}-\boldsymbol{k'}|^3}\Pi_\lambda^{ij}(\boldsymbol{k})\Pi_{\lambda'}^{pq}(\boldsymbol{k})^* \nonumber \\
    &\times\sum\limits_{\lambda_1,\lambda_2=\pm}\Bigg[(\boldsymbol{k}.\boldsymbol{k'}-k'^2)^2\varepsilon_{m i}^{(\lambda_1)}(\boldsymbol{k'})\varepsilon_{\alpha p}^{(\lambda_1)}(\boldsymbol{k'})^*\varepsilon_{j\ell}^{(\lambda_2)}(\boldsymbol{k}-\boldsymbol{k'})\varepsilon_{\alpha q}^{(\lambda_2)}(\boldsymbol{k}-\boldsymbol{k'})^* \nonumber \\
    &+(\boldsymbol{k}.\boldsymbol{k'}-k'^2)k'_\alpha(k_\beta-k'_\beta)\varepsilon_{m i}^{(\lambda_1)}(\boldsymbol{k'})\varepsilon_{\beta p}^{(\lambda_1)}(\boldsymbol{k'})^*\varepsilon_{mj}^{(\lambda_2)}(\boldsymbol{k}-\boldsymbol{k'})\varepsilon_{q\alpha}^{(\lambda_2)}(\boldsymbol{k}-\boldsymbol{k'})^* \nonumber \\
    &+(\boldsymbol{k}.\boldsymbol{k'}-k'^2)k'_\ell(k_m-k'_m)\varepsilon_{m i}^{(\lambda_1)}(\boldsymbol{k'})\varepsilon_{\beta p}^{(\lambda_1)}(\boldsymbol{k'})^*\varepsilon_{j\ell}^{(\lambda_2)}(\boldsymbol{k}-\boldsymbol{k'})\varepsilon_{\beta q}^{(\lambda_2)}(\boldsymbol{k}-\boldsymbol{k'})^* \nonumber \\
    &+k'_\ell(k_m-k'_m)k'_\alpha(k_\beta-k'_\beta)\varepsilon_{m i}^{(\lambda_1)}(\boldsymbol{k'})\varepsilon_{\beta p}^{(\lambda_1)}(\boldsymbol{k'})^*\varepsilon_{j\ell}^{(\lambda_2)}(\boldsymbol{k}-\boldsymbol{k'})\varepsilon_{q\alpha}^{(\lambda_2)}(\boldsymbol{k}-\boldsymbol{k'})^*\Bigg] \nonumber \\
    &\times\Bigg\lvert\int\limits_{\eta_{\rm in}}^{\eta_{\rm end}}\dfrac{d\eta_1}{a(\eta_1)^2}G_k(\eta_{\rm end},\eta_1)\xi_Z(k',|\boldsymbol{k}-\boldsymbol{k'}|,\eta_1)\Bigg\lvert^2\:,
\end{align}
\end{widetext}
where $\Re(x)$ denotes the real part of $x$. In the expressions above, we have defined $f_Z(k_1,k_2,\eta)=T_{k_1}'(\eta)T_{k_2}'(\eta)$ and $\xi_Z(k_1,k_2,\eta)=T_{k_1}(\eta)T_{k_2}(\eta)$ based on the inflationary transfer function $T_k(\eta)$ introduced in \eqref{eq:inftransf}. The resulting GW power spectrum $\mathcal{P}_h^{(Z)}(k,\eta_{\rm end})$ can then be constructed following \eqref{eq:tenspowspec}, and the GW spectral abundance at the present time may be calculated as $\Omega_{\rm GW}(k,\eta_0)=T_h\left(k;\eta_0,\eta_{\rm end}\right)^2\mathcal{P}_h^{(Z)}(k,\eta_{\rm end})$, where the post-inflationary sub-horizon tensor transfer function (squared and oscillation-averaged) is given by \cite{Watanabe:2006qe}
\begin{widetext}
\begin{equation}
    T_h\left(k;\eta_0,\eta_{\rm end}\right)^2=
    \begin{cases}
        \dfrac{1}{12}\left(\dfrac{k}{H_0}\right)^2\left(\dfrac{\eta_{\rm eq}}{\eta_0}\right)^2\left[A(k)j_2(k\eta_0)+B(k)y_2(k\eta_0)\right]^2\::\:\:k>k_{\rm eq}\:, \\
        \dfrac{1}{12}\left(\dfrac{k}{H_0}\right)^2\left[\dfrac{3j_2(k\eta_0)}{k\eta_0}\right]^2\::\:\:k<k_{\rm eq}\:,
    \end{cases}
\end{equation}
\end{widetext}
where the coefficients $A(k)$ and $B(k)$ are
\begin{equation}
    A(k)=\dfrac{3}{2k\eta_{\rm eq}}-\dfrac{\cos(2k\eta_{\rm eq})}{2k\eta_{\rm eq}}+\dfrac{\sin(2k\eta_{\rm eq})}{(k\eta_{\rm eq})^2}\:,
\end{equation}
\begin{equation}
    B(k)=-1+\dfrac{1}{(k\eta_{\rm eq})^2}-\dfrac{\cos(2k\eta_{\rm eq})}{(k\eta_{\rm eq})^2}-\dfrac{\sin(2k\eta_{\rm eq})}{2k\eta_{\rm eq}}\:.
\end{equation}


\bibliography{biblio}

\end{document}